\documentstyle[twocolumn,aps,epsf]{revtex}

\begin{document}

\draft

\title{Evaporative Cooling of a Two-Component Degenerate Fermi Gas}

\author{M. J. Holland, B. DeMarco, and D. S. Jin}

\address{JILA, University of Colorado and National Institute of
  Standards and Technology, Boulder, Colorado 80309-0440}

\date{\today}

\wideabs{

\maketitle

\begin{abstract}
  We derive a quantum theory of evaporative cooling for a degenerate
  Fermi gas with two constituents and show that the optimum cooling
  trajectory is influenced significantly by the quantum statistics of
  the particles. The cooling efficiency is reduced at low temperatures
  due to Pauli blocking of available final states in each binary
  collision event. We compare the theoretical optimum trajectory with
  experimental data on cooling a quantum degenerate cloud of
  potassium-40, and show that temperatures as low as 0.3 times the
  Fermi temperature can now be achieved.
\end{abstract}


}

The recent demonstrations of Bose-Einstein condensation in dilute
alkali and hydrogen gases have required the ability to reach extremely
low temperatures in the micro-Kelvin to nano-Kelvin scale.  To date,
this has been possible only by using the experimental technique of
forced evaporative cooling~\cite{hess}. Efficient evaporative cooling
can allow the temperature of a gas to be reduced by orders of
magnitude without prohibitive loss in the number of atoms.  It has
universal application to cool magnetically trapped atomic and
molecular vapors and has already been applied to produce quantum
degenerate clouds of rubidium~\cite{rubidium}, sodium~\cite{sodium},
lithium~\cite{lithium}, potassium~\cite{potassium}, and
hydrogen~\cite{hydrogen}.

For a bosonic gas, cooling can be continued to the point where no
discernible normal component of the gas is present, closely
approximating a zero temperature system.  Demonstrating the ability to
reach this regime has been a prerequisite to many of the recent
experiments on collective effects in these systems. Collective
phenomena that have now been observed include linear
response~\cite{zerotemp,finitetemp,collective,hydro}, surface
modes~\cite{surface}, and topological excitations such as
vortices~\cite{vortexth,vortexex}. A current goal is to observe the
conjugate low-temperature phenomena in a fermionic gas when it is
cooled well below the onset of quantum
degeneracy~\cite{amoruso,vichi,bruun,yip}.

In evaporative cooling, a ``cut'' is made at a prescribed energy and
all atoms with energies greater than the cut are removed from the
system. The remaining atoms will rethermalize by collisions to form an
equilibrium with a lower temperature. The crucial parameter that
determines the timescale for cooling is therefore the rate of
rethermalization $\Gamma$. For a dilute gas at temperatures where
quantum statistics do not play a role, rethermalization is determined
by the elastic collision rate, given by $\Gamma=\bar n\sigma v$, where
$\bar n$ is the spatially averaged density-weighted density, $\sigma$
is the collision cross section, and $v$ is the root-mean-square
velocity of the colliding species. In a harmonic trap, $\Gamma$ may
increase as the gas is cooled despite the obvious reduction in average
velocity.  The reason for this is simply that as the cloud cools, the
atoms fall to the bottom of the trap and become more tightly confined,
increasing the number density $n$ and more than compensating for the
loss in energy per particle. Achieving this regime, known as {\em
  runaway evaporation}, is typically an experimental prerequisite for
following a cooling trajectory that leads to a quantum degenerate gas.

For bosons, once the temperature is reduced below the critical
temperature for Bose-Einstein condensation to occur, effects due to
quantum statistics assist the evaporation. Consider a typical
collision event involving two atoms from the normal thermal gas that
initially have approximately the mean energy of the distribution. The
presence of a Bose-Einstein condensate modifies the scattering
probability into each possible final state and enhances the likelihood
of stimulated scattering of one of the atoms into the condensate. The
other atom then obtains the total energy of the initial pair and can
be removed by the evaporative cut. Clearly this type of collision
leads to very efficient evaporative cooling.

The opposite situation is true for fermions. As the temperature falls
below the Fermi temperature, efficient collisions turn off due to
Pauli blocking since the states of lowest energy become occupied with
high probability~\cite{ferrari}. In this paper we study this effect on
the achievable optimum evaporation trajectory.  Our calculations are
motivated by the first application of evaporative cooling to produce a
quantum degenerate Fermi gas~\cite{potassium}. This recent experiment
has opened the door to the study of Fermi statistics in an extremely
dilute regime---perhaps eventually allowing for the possibility of
investigating Cooper pairing and the BCS phase transition in these
dilute systems.

At typical temperatures of interest, collisions between atom pairs are
purely $s$-wave since the characteristic collision energies are well
below the centrifugal barrier associated with channels of non-zero
orbital angular momenta~\cite{thresholdlaw}. Since for fermions the
total wavefunction must be antisymmetric with respect to exchange of
any pair of atoms, $s$-wave collisions are only possible if at least
two internal atomic hyperfine states are simultaneously present, or
alternatively if sympathetic cooling is performed with a
distinguishable species, such as a different isotope or a different
element~\cite{wiemanexpt,liyou}.  Here, we consider the first of these
possibilities---a two-component Fermi gas.

The Hamiltonian for this system may be separated into two parts,
${H}={H}_0+{H}_1$ where ${H}_0$ is the usual single particle energy of
the system and ${H}_1$ describes binary collisions. The Hamiltonian
${H}_0$ for a two-component mixture confined in a three-dimensional
harmonic oscillator is
\begin{equation}
  {H}_0=\sum_{\mbox{\boldmath$\scriptstyle n$}}
  E_{\mbox{\boldmath$\scriptstyle n$}}
  \left(a^{\dag}_{\mbox{\boldmath$\scriptstyle n$}}
    a^{\phantom\dag}_{\mbox{\boldmath$\scriptstyle n$}} +
    b^{\dag}_{\mbox{\boldmath$\scriptstyle n$}}
    b^{\phantom\dag}_{\mbox{\boldmath$\scriptstyle n$}} \right),
\end{equation}
where we have assumed the potential is identical for both species. The
summation is taken over the three integer components of
$\mbox{\boldmath$n$}=(n_x,n_y,n_z)$. If the harmonic potential is
isotropic with oscillation frequency~$\omega$ then
$E_{\mbox{\boldmath$\scriptstyle n$}} = \hbar\omega(n_x+n_y+n_z+3/2)$.
The annihilation operators for the two components,
$a_{\mbox{\boldmath$\scriptstyle n$}}$ and
$b_{\mbox{\boldmath$\scriptstyle n$}}$, obey the usual Fermi
commutation relations. Binary collisions are described by the
interaction Hamiltonian
\begin{equation}
  \hat{H}_1=\sum_{ \mbox{\boldmath$\scriptstyle n$},
    \mbox{\boldmath$\scriptstyle m$}, \mbox{\boldmath$\scriptstyle
      q$}, \mbox{\boldmath$\scriptstyle p$}}
  C_{\mbox{\boldmath$\scriptstyle n$}, \mbox{\boldmath$\scriptstyle
      m$}; \mbox{\boldmath$\scriptstyle q$},
    \mbox{\boldmath$\scriptstyle p$}}\,\,
  a^{\dag}_{\mbox{\boldmath$\scriptstyle n$}}
  b^{\dag}_{\mbox{\boldmath$\scriptstyle m$}}
  b^{\phantom\dag}_{\mbox{\boldmath$\scriptstyle q$}}
  a^{\phantom\dag}_{\mbox{\boldmath$\scriptstyle p$}},
\end{equation}
where the matrix element is
\begin{equation}
  C_{\mbox{\boldmath$\scriptstyle n$}, \mbox{\boldmath$\scriptstyle
      m$}; \mbox{\boldmath$\scriptstyle q$},
    \mbox{\boldmath$\scriptstyle p$}} = U_0\int d^3x\,
  {\alpha}^*_{\mbox{\boldmath$\scriptstyle n$}}(\mbox{\boldmath$x$})
  {\beta}^*_{\mbox{\boldmath$\scriptstyle m$}}(\mbox{\boldmath$x$})
  {\beta}^{\phantom*}_{\mbox{\boldmath$\scriptstyle
      q$}}(\mbox{\boldmath$x$})
  {\alpha}^{\phantom*}_{\mbox{\boldmath$\scriptstyle
      p$}}(\mbox{\boldmath$x$}).
\end{equation}
The oscillator eigenfunctions ${\alpha}_{\mbox{\boldmath$\scriptstyle
    n$}}$ and ${\beta}_{\mbox{\boldmath$\scriptstyle n$}}$ form a
complete orthonormal basis that spans the two-component Hilbert space.
In calculating the matrix element, we have replaced the physical
two-particle potential by a contact potential. The dimensional
prefactor is $U_0=4\pi\hbar^2 a/m$, where $m$ is the atomic mass, and
$a$ is the $s$-wave scattering length, which includes contributions
from both direct and exchange scattering.

Although, in principle, one could solve the evolution of this isolated
many-body system, we are primarily interested here in a simplified
description on a coarse-grained timescale. Such a description is given
by quantum kinetic theory where a set of relevant observables are
quantities of interest. The underlying theoretical framework derives
from the property that collisions in the dilute gas are extremely well
separated in time. This allows the Born and Markov approximations to
be made and the subsequent derivation of a perturbative theory in
lowest orders of the interaction Hamiltonian~\cite{kadanoff}.  The
relevant observables are the populations of the two species (diagonal
elements of the single particle density matrix):
$A_{\mbox{\boldmath$\scriptstyle n$}}=\langle
a^{\dag}_{\mbox{\boldmath$\scriptstyle n$}}
a^{\phantom\dag}_{\mbox{\boldmath$\scriptstyle n$}} \rangle$ and
$B_{\mbox{\boldmath$\scriptstyle n$}}=\langle
b^{\dag}_{\mbox{\boldmath$\scriptstyle n$}}
b^{\phantom\dag}_{\mbox{\boldmath$\scriptstyle n$}}\rangle$.

This approach must be extended in order to treat boson-fermion
mixtures, or Fermi gases at temperatures where Cooper pairing is
important. In those situations it would be necessary to expand the set
of relevant observables to consider diagonal and off-diagonal
contributions to the normal and anomalous densities, as well as the
role of mean-fields~\cite{kineticpaper}.

Following this procedure, the quantum kinetic equations (also referred
to as the ``Quantum Boltzmann Equations'') for the two-component Fermi
gas are given by
\begin{eqnarray}
  {dA_{\mbox{\boldmath$\scriptstyle n$}}\over dt} &=& {1\over2} \sum_{
    \mbox{\boldmath$\scriptstyle m$}, \mbox{\boldmath$\scriptstyle
      q$}, \mbox{\boldmath$\scriptstyle p$}}
  W_{\mbox{\boldmath$\scriptstyle n$}, \mbox{\boldmath$\scriptstyle
      m$}; \mbox{\boldmath$\scriptstyle q$},
    \mbox{\boldmath$\scriptstyle p$}}\,
  \Bigl\{A_{\mbox{\boldmath$\scriptstyle
      p$}}B_{\mbox{\boldmath$\scriptstyle
      q$}}(1-B_{\mbox{\boldmath$\scriptstyle m$}})
  (1-A_{\mbox{\boldmath$\scriptstyle
      n$}})\nonumber\\
  &&\quad {} - A_{\mbox{\boldmath$\scriptstyle
      n$}}B_{\mbox{\boldmath$\scriptstyle
      m$}}(1-B_{\mbox{\boldmath$\scriptstyle
      q$}})(1-A_{\mbox{\boldmath$\scriptstyle p$}})\Bigr\},\nonumber\\
  {dB_{\mbox{\boldmath$\scriptstyle m$}}\over dt} &=& {1\over2} \sum_{
    \mbox{\boldmath$\scriptstyle n$}, \mbox{\boldmath$\scriptstyle
      q$}, \mbox{\boldmath$\scriptstyle p$}}
  W_{\mbox{\boldmath$\scriptstyle n$}, \mbox{\boldmath$\scriptstyle
      m$}; \mbox{\boldmath$\scriptstyle q$},
    \mbox{\boldmath$\scriptstyle p$}}\,
  \Bigl\{A_{\mbox{\boldmath$\scriptstyle
      p$}}B_{\mbox{\boldmath$\scriptstyle
      q$}}(1-B_{\mbox{\boldmath$\scriptstyle
      m$}})(1-A_{\mbox{\boldmath$\scriptstyle n$}})\nonumber\\
  &&\quad {} - A_{\mbox{\boldmath$\scriptstyle
      n$}}B_{\mbox{\boldmath$\scriptstyle
      m$}}(1-B_{\mbox{\boldmath$\scriptstyle
      q$}})(1-A_{\mbox{\boldmath$\scriptstyle p$}})\Bigr\},
\label{qk1}
\end{eqnarray}
where the transition rates are found from Fermi's golden rule
\begin{equation}
  W_{\mbox{\boldmath$\scriptstyle n$}, \mbox{\boldmath$\scriptstyle
      m$}; \mbox{\boldmath$\scriptstyle q$},
    \mbox{\boldmath$\scriptstyle p$}} = {2\pi\over\hbar}
  |C_{\mbox{\boldmath$\scriptstyle n$}, \mbox{\boldmath$\scriptstyle
      m$}; \mbox{\boldmath$\scriptstyle q$},
    \mbox{\boldmath$\scriptstyle
      p$}}|^2{\delta_{E_{\mbox{\boldmath$\scriptscriptstyle
          m$}}+E_{\mbox{\boldmath$\scriptscriptstyle
          n$}},E_{\mbox{\boldmath$\scriptscriptstyle
          q$}}+E_{\mbox{\boldmath$\scriptscriptstyle p$}}}
    \over \hbar\omega}.
\label{collkern}
\end{equation}
Note the factors such as $(1-A_{\mbox{\boldmath$\scriptstyle n$}})$
and $(1-B_{\mbox{\boldmath$\scriptstyle m$}})$ give rise to the
mechanism known as Pauli blocking. The Kronecker delta function,
$\delta$, constrains the collision to be precisely on the energy
shell. This is an unphysical artifact of the Markov approximation that
is valid for the dilute gas only when the elastic collision rate is
much less than the oscillation frequency in the trap. Otherwise the
collisional broadening of the levels would be greater than their
spacing, and the energy basis we use here would not be an appropriate
choice since off-diagonal elements would then be important. We assume
ergodicity by assigning equal population to each state in the
degenerate manifold of states with the same principal quantum number
$n$ (and therefore the same energy). We denote the ergodic populations
by $A_{e_n}$ and $B_{e_n}$ (indexed only by the discrete values of the
energy $e_n$) which are related to previously defined populations for
an arbitrary quantum state $A_{\mbox{\boldmath$\scriptstyle n$}}$ and
$B_{\mbox{\boldmath$\scriptstyle n$}}$ by
\begin{eqnarray}
  g_{e_n}A_{e_n} &=& \sum_{\mbox{\boldmath$\scriptstyle
      n$}}\delta_{e_n,E_{\mbox{\boldmath$\scriptscriptstyle
        n$}}}A_{\mbox{\boldmath$\scriptstyle n$}},\nonumber\\
  g_{e_n}B_{e_n} &=& \sum_{\mbox{\boldmath$\scriptstyle
      n$}}\delta_{e_n,E_{\mbox{\boldmath$\scriptscriptstyle
        n$}}}B_{\mbox{\boldmath$\scriptstyle n$}},
\end{eqnarray}
with $g_{e_n}$ denoting the degeneracy of states for the
three-dimensional harmonic oscillator
\begin{equation}
  g_{e_n}={1\over2}(n+1)(n+2).
\label{degeneracy}
\end{equation}
The quantum kinetic equations given in Eq.~(\ref{qk1}) may be
simplified by approximating the summations by integrals over
continuous distributions. This is typically always a good
approximation for fermions, since for a sufficiently large sample, a
macroscopic number of states are occupied even at very low
temperature. The quantum kinetic equations then describe the rate of
transfer of population between continuous distribution functions.  We
denote these functions (of a continuous energy variable $e$) as $A(e)$
and $B(e)$, which evolve according to
\begin{eqnarray}
  \rho(e_n){dA(e_n)\over dt} &=& {m\sigma\over\pi^2\hbar^3} \int
  de_mde_q
  de_p\, \delta(\Delta)\rho(e_{\rm min})\nonumber\\
  &&\Bigl\{A(e_p)B(e_q)[1-B(e_m)][1-A(e_n)]\nonumber\\
  &&-A(e_n)B(e_m)[1-B(e_q)][1-A(e_p)]\Bigr\}\nonumber\\
  \rho(e_m){dB(e_m)\over dt} &=& {m\sigma\over\pi^2\hbar^3} \int
  de_nde_q
  de_p\, \delta(\Delta)\rho(e_{\rm min})\nonumber\\
  &&\Bigl\{A(e_p)B(e_q)[1-B(e_m)][1-A(e_n)]\nonumber\\
  &&-A(e_n)B(e_m)[1-B(e_q)][1-A(e_p)]\Bigr\},
\label{qk2}
\end{eqnarray}
where $e_{\rm min}=\min\{e_n,e_m,e_p,e_q\}$ and
$\Delta=e_n+e_m-e_q-e_p$. The quantum mechanical cross-section
applicable here is $\sigma=4\pi a^2$ since the products of a collision
event are in quantum mechanically distinguishable spin
states~\cite{burke}. The density of states $\rho(e)$ for the
three-dimensional harmonic oscillator can be found from the large $n$
limit of Eq.~(\ref{degeneracy}), which gives $\rho(e)={1\over2}e^2$.
Although these equations are similar in form to the discrete version
in Eq.~(\ref{qk1}), a key simplification has been the replacement of
the collision kernel [defined in Eq.~(\ref{collkern})] by $\rho(e_{\rm
  min})$; which is the classical limit~\cite{walraven}. As was shown
in Ref.~\cite{simtraj}, the convergence to the classical limit is very
rapid as $e_{\rm min}$ is raised.  Significant quantum correction
occurs only when both of the colliding atoms are in the lowest few (of
order one to five) states of the harmonic oscillator.  These lowest
energy collisions give a microscopic correction to the collision rate
when the particles have Fermi statistics and are typically distributed
over a macroscopic number of levels of the oscillator. The situation
is quite different for the quantum degenerate Bose gas where a careful
treatment of these low energy collisions is usually crucial due to the
possibility of condensate mean-fields.

\begin{figure}
\begin{center}\
\epsfysize=60mm
\epsfbox{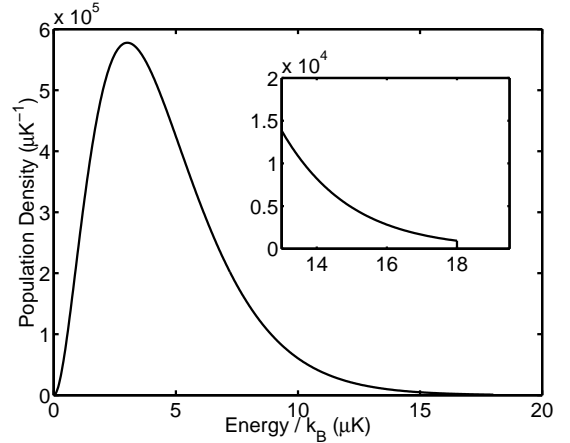}
\end{center}
\caption{The truncated Fermi-Dirac distribution. The graph shows the 
  distribution function, $\rho(e)F(e)$, as a function of $e/k_B$ where
  $k_B$ is Boltzmann's constant. The parameters used were those for
  the start of the evaporation simulation with $N=3.2\times10^6$,
  $E=3Nk_BT$ where $T=1.5\,\mu$K, and $K=4E/N$. The inset shows a
  magnified view of the discontinuity at the cut energy.}
\label{trunfd}
\end{figure}

While the simultaneous equations in Eq.~(\ref{qk2}) may be solved by
direct numerical integration, the calculation is cumbersome given the
multidimensional integrals that must be performed at each timestep. We
provide a more intuitive approach that is motivated by the near
equilibrium distributions expected when the elastic collision rate is
sufficiently high. We assume that the form of the population
distribution functions for both components are given by a truncated
Fermi-Dirac distribution $F(e)$ as defined by
\begin{equation}
  F(e)=\left\{
\begin{array}{cc}
(\exp[{\beta(e-\mu)}]+1)^{-1}  & e<K \\
0 & {\rm otherwise} 
\end{array}
\right..
\label{dist}
\end{equation}
A similar method was introduced in Ref.~\cite{walraven} to treat the
evaporative cooling of a classical gas. For simplicity, we also take
the simplest case of $A(e)=B(e)=F(e)$ at all times, since if the
distributions of the two components are initially identical they will
remain so due to the symmetry of the equations.  This means that the
distribution functions for both species are parameterized by the same
three variables: (i) an inverse temperature $\beta$, (ii) a chemical
potential $\mu$, and (iii) a cut energy $K$. Given a value for the cut
energy, $\beta$ and $\mu$ can be solved to give simultaneously the
correct total number of atoms in one component $N$, and total energy
of these atoms $E$, according to
\begin{eqnarray}
  \int_0^K de\,\rho(e)F(e)=N,\nonumber\\
  \int_0^K de\, e\rho(e)F(e)=E.
\label{constrain}
\end{eqnarray}
The truncated Fermi-Dirac distribution function is illustrated in
Fig.~\ref{trunfd} for the initial conditions of the evaporative
simulation.

The simulation algorithm is the following:
\begin{enumerate}
\item Starting with given $N$, $E$, and $K$, solve
  Eq.~(\ref{constrain}) to find $\beta$ and $\mu$.
\item Consider propagation of the kinetic equations for a time step
  $dt$ and determine the change in number $dN_1$ and energy $dE_1$ due
  to atoms colliding and gaining energy above the cut $K$,
\begin{eqnarray}
  dN_1&=&dt{m\sigma\over\pi^2\hbar^3}
  \int_K^{2K}de_n\int_0^{2K-e_n}de_m \int_{e_m+e_n-K}^Kde_p
  \nonumber\\
  &&\rho(e_m)F(e_m+e_n-e_p)F(e_p)[1-F(e_m)]\nonumber\\
  dE_1&=&dt{m\sigma\over\pi^2\hbar^3}
  \int_K^{2K}de_n\int_0^{2K-e_n}de_m \int_{e_m+e_n-K}^Kde_p
  \nonumber\\
  &&e_n\rho(e_m)F(e_m+e_n-e_p)F(e_p)[1-F(e_m)]
\end{eqnarray}
\item Simulate background loss (energy-independent removal of atoms
  arising from non-ideal vacuum conditions in experiments) with rate
  $\gamma$ from the trap
\begin{eqnarray}
  dN_{2}&=&\gamma Ndt,\nonumber\\
  dE_{2}&=&\gamma Edt.
\end{eqnarray}
\item Lower the cut energy, from $K$ to a new value $K'$ and find the
  change in number and energy due to trimming the highest energy atoms:
\begin{eqnarray}
  dN_{3}&=&\int_{K'}^Kde\,\rho(e)F(e),\nonumber\\
  dE_{3}&=&\int_{K'}^Kde\,e\rho(e)F(e).
\end{eqnarray}
\item Update the number and energy
\begin{eqnarray}
  N&\rightarrow& N-\sum_{\sigma} dN_{\sigma}, \nonumber\\
  E&\rightarrow& E-\sum_{\sigma} dE_{\sigma},
\end{eqnarray}
and repeat this sequence starting again from step~1.
\end{enumerate}

A technical point is that solving Eq.~(\ref{constrain}) for $\beta$
and $\mu$ is a two-dimensional root finding problem that can
potentially be non-trivial. We use a multidimensional Newton-Raphson
algorithm that will rapidly produce a good estimate of the value of
the solutions in a few iterations. To find a good estimate to start
with, we employ a simple three-point polynomial extrapolation of
solutions from previous timesteps. In this extrapolation we use the
cut energy $K$ as the independent variable, rather than the step
number or time.

Although this method may be used to calculate the evaporation
trajectory for an arbitrary time dependence of the cut energy, we are
most interested here in determining the optimum path. During the
evaporation simulation, we follow a trajectory that maximizes the
energy removed per particle from the system. That is, we choose a
value for $K'$ in such a way as to numerically maximize
\begin{equation}
  {\sum_{\sigma}dE_{\sigma}\over \sum_{\sigma}dN_{\sigma}}
\end{equation}
for the subsequent timestep.

\begin{figure}
\begin{center}\
  \epsfysize=130mm \epsfbox{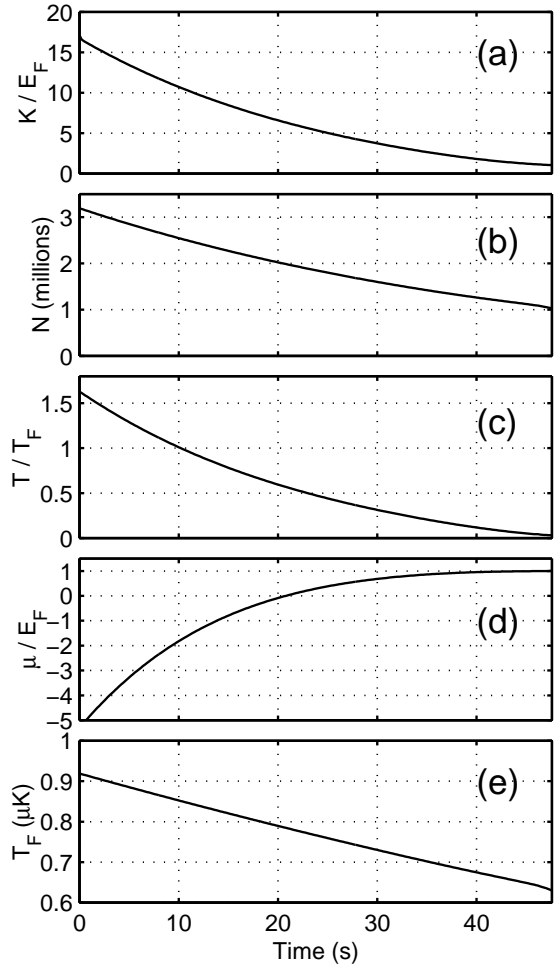}
\end{center}
\caption{Optimum evaporation trajectory. The parameters used 
  were those typical for the experiment reported in
  Ref.~\protect\cite{potassium}; $\omega/2\pi=70$~Hz is the geometric
  mean of the trap frequencies, $m$ is the mass of potassium-40,
  $\gamma=1/350$~Hz, and the scattering is dominated by the triplet
  channel with $a=157a_0$ where $a_0$ is the Bohr
  radius~\protect\cite{thresholdlaw}. The initial conditions used are
  given in Fig.~\ref{trunfd}. (a) The optimized cut energy $K$ divided
  by the Fermi energy $E_F$. For a three-dimensional harmonic
  oscillator the Fermi energy is $E_F=(6N)^{1/3}\hbar\omega$.  (b) The
  number of atoms in each component. (c) The temperature $T$, defined
  for the truncated Fermi-Dirac distribution as $T=1/(\beta k_B)$. The
  temperature is normalized by the Fermi temperature $T_F=E_F/k_B$.
  (d) The chemical potential $\mu$ in units of $E_F$.  (e) The Fermi
  temperature $T_F$.}
\label{simulation}
\end{figure}

In Fig.~\ref{simulation}, we show the calculated optimum evaporation
trajectory for the two-component Fermi gas. In order to indicate the
level of quantum degeneracy, we have normalized energies and
temperatures by dividing them by the Fermi energy and Fermi
temperature (see caption). The optimum cut energy approaches closely
the Fermi surface towards the end of the simulation. While the ideal
evaporative trajectory demonstrates the theoretical possibility for
achieving very low temperatures---with the chemical potential tending
towards the Fermi energy and with a macroscopic population
remaining---the efficiency of the evaporation trajectory falls
dramatically as the system becomes degenerate. 

\begin{figure}
\begin{center}\
  \epsfysize=60mm \epsfbox{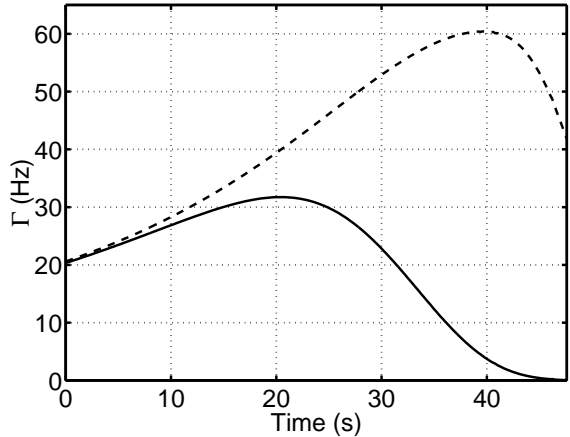}
\end{center}
\caption{Elastic collision rate per atom for the optimized 
  evaporation trajectory. The solid line shows the true collision rate
  as defined in Eq.~(\ref{collrate}). The dashed line shows what the
  collision rate would be in the absence of Pauli blocking, by
  artificially replacing the $[1-F(e_m)][1-F(e_n)]$ factors in
  Eq.~(\ref{collrate}) by unity.}
\label{collratefig}
\end{figure}

This is shown by the elastic collision rate $\Gamma$ defined by
\begin{eqnarray}
  \Gamma&=&{m\sigma\over2N\pi^2\hbar^3} \int
  de_nde_mde_pde_q\,\delta(\Delta)\rho(e_{\rm
    min}) \nonumber\\
  &&\quad F(e_p)F(e_q)[1-F(e_m)][1-F(e_n)],
\label{collrate}
\end{eqnarray}
which is illustrated in Fig.~\ref{collratefig}. As the chemical
potential becomes positive, and Pauli blocking of available final
states begins to play an important role, the elastic collision rate
falls sharply. Since the elastic collision rate determines the
timescale for rethermalization, at the end of the simulation,
evaporative cooling has virtually ceased. As this figure dramatically
illustrates, towards the end of the evaporation trajectory, the
elastic collision rate may be more than an order of magnitude
suppressed from the value it would have if Pauli blocking of final
states was absent.

\begin{figure}
\begin{center}\
  \epsfysize=60mm \epsfbox{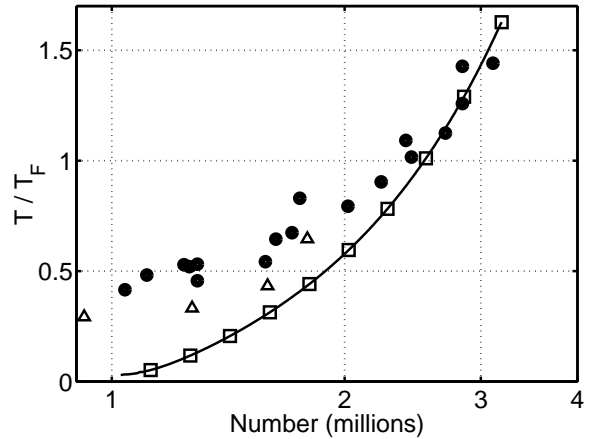}
\end{center}
\caption{Optimum evaporation trajectory. The $\Box$ symbols 
  show points on the theoretical evaporation trajectory at time
  intervals of 5~s. For comparison, the $\bullet$ symbols show
  experimental data points giving a typical evaporation trajectory.
  The $\triangle$ symbols show the lowest $T/T_F$ achieved in our
  current experiment, as described in the text.}
\label{phasespace}
\end{figure}

In Fig.~\ref{phasespace}, we illustrate this trajectory on a semilog
graph of temperature versus number and compare with experimental data.
The data are taken by evaporating a two-component gas of $^{40}$K as
described in Ref.~\cite{potassium}. For the portion of the evaporation
trajectory shown in Fig.~\ref{phasespace} evaporation occurs using a
50/50 mixture of two spin states confined in a cylindrically symmetric
harmonic trap whose radial frequency is 137~Hz and axial frequency is
19.5~Hz. After evaporation, one of the spin components is removed
quickly (within 0.3~s) with the application of a frequency-swept
microwave field; this removal provides a small amount of additional
evaporative cooling that reduces the cloud temperature by 20\%. The
comparison shown in Fig.~\ref{phasespace} illustrates that although it
is possible theoretically to reach very low temperatures, the data
corresponds to less efficient evaporative cooling and is presumably
limited by experimental artifacts. These experimental limitations
could include heating of the trapped gas, finite energy resolution of
the evaporative cut, reduced dimensionality of the evaporative cut,
and other similar problems. 

In the current experiment we have recently found that improving the
stability of the magnetic trapping field increased the highest
achievable quantum degeneracy from $T/T_F=0.5$ to $T/T_F=0.3$. The low
temperature part of an experimental exaporation trajectory that
reached $T/T_F=0.3$ is also shown in Fig.~\ref{phasespace}. For this
data we used a much slower removal of the second spin component
(within 25~s) to provide additional evaporative cooling which is not
included in the theory. The experimental progress suggests that
further technical improvements may enable experiments to approach the
low $T/T_F$ values that appear possible theoretically. Furthermore
Fig.~\ref{collratefig} shows that the dramatic suppression of the
elastic collision rate due to Pauli blocking could be observed at the
lowest temperatures of current experiments.

We would like to thank J.~Cooper, R.~Walser, B. Anderson, and J.~Bohn
for discussions. M.~Holland acknowledges support for this work from
the Department of Energy.  D.~Jin and B.~DeMarco acknowledge funding
from the Office of Naval Research and the National Science Foundation.

\end{document}